\documentclass{aa}
\usepackage{graphics}

\begin{document}

\thesaurus{11(11.17.2; 11.17.3; 11.17.4)}

\title{Spectroscopy of the Extended Emission Associated with
Two High-z Quasars\thanks{Based on observations taken at the
W. M. Keck Observatory}}

\author{Matthew D. Lehnert\inst{1,2}, \and Robert H. Becker \inst{2,3}}
\institute{Sterrewacht Leiden, P.O. Box 9513, 2300 RA, Leiden, The Netherlands
\and Institute of Geophysics and Planetary Physics, Lawrence Livermore
National Laboratory, 7000 East Ave., L-413, Livermore, CA 94550 \and
Department of Physics and Astronomy, University of California Davis}

\date{Received 23 June 1997 / Accepted 5 Nov 1997}

\titlerunning{Extended Emission Associated with Two High-z Quasars}
\authorrunning{Lehnert \& Becker}
\offprints{M. Lehnert}

\maketitle

\begin{abstract}

We present spectra taken with the Low Resolution Imaging Spectrograph
on the Keck 10m telescope of spatially-resolved structures (\lq
fuzz\rq) around the high-redshift radio-loud quasars \object{PKS
0445+097} (z=2.108) and \object{PKS 2338+042} (z=2.589).  For
\object{PKS 0445+097} we oriented the slit such that it passed through
the quasar nucleus and through an object identified on an HST exposure
1.8'' to the southeast of the nucleus.  We find that this object has
strong [OII]$\lambda$3727 emission and weaker emission due to
[NeV]$\lambda$3426 and [NeIII]$\lambda$3869, at a redshift of
z=0.8384.  This redshift is very close ($\Delta$v=200 km s$^{-1}$) to
that of a MgII absorption line identified in the spectrum of
\object{PKS 0445+097}.  The line ratios and narrowness of the lines
indicate that this galaxy probably has a Seyfert 2 nucleus.  Moreover,
we find that there is faint extended HeII$\lambda$1640 and
CIII]$\lambda$1909 from the quasar host galaxy. Both of the lines are
broad (FWHM$_{HeII}$=1000 $\pm$ 200 km s$^{-1}$ and FWHM=2200 $\pm$ 600
kms$^{-1}$).  The limits on the fluxes, the large line widths, and the
line ratio HeII/CIII] are similar to that observed for high redshift
radio galaxies.

From the spectrum of \object{PKS 2338+042}, we find extended Ly$\alpha$ emission
on scales about 10'' which is strongly one sided (on the side of the
stronger, closer to the nucleus, more distorted radio lobe).  The
extended Ly$\alpha$ line is broad with widths of over 1000 km
s$^{-1}$.  The extended emission shows no evidence for a strong
velocity shear.  We also find weak extended CIV and HeII emission which
has similar characteristics (line widths, line fluxes, and line ratios)
to that of high redshift radio galaxies.  In addition, we
serendipitously discovered what appears to be a Ly$\alpha$ emitting
galaxy about 29'' to the southeast of the quasar at z=2.665.  The
single line identified in the spectrum has a measure equivalent width
of 2100\AA \ and a full width at half maximum of 2150 km s$^{-1}$.  The
high equivalent width and large velocity width of the putative
Ly$\alpha$ line suggest that this galaxy must be a high redshift AGN,
although we note that the upper limits on the CIV and HeII line ratios
are mildly inconsistent with this galaxy harboring an AGN.  An
approximate calculation shows the quasar nucleus is not exciting the
Ly$\alpha$ emission from this galaxy.

\end{abstract}

\section{Introduction}

Luminous active galactic nuclei (AGN) are very much a high-redshift
phenomenon -- their co-moving space density has fallen by about a
factor of 1000 between the epochs z$\approx$2.5 and the present (e.g.,
\cite{hs90}; \cite{boyle}).  Many current cosmogonical scenarios suggest
that the era of major star-formation within galaxy spheroids (the
``epoch of galaxy formation'') took place at about the same time as
this dramatic peak in the co-moving space densities of quasars (e.g.,
\cite{rees88}).  If this is indeed the case, then by gaining insight into
the processes which govern the creation, evolution, and ``mass
extinction'' of AGN, we can perhaps achieve a deeper understanding of
the mechanisms that control the evolution of galaxies in general.

Recently, there has been a breakthrough in our understanding of the
environments of radio-loud quasars at z$\approx$2. Ground-based data
and more recently obtained high-quality HST imaging data have revealed
many fascinating properties of the host galaxies of high redshift
quasars.  It has been found that quasar fuzz has strong nebular
emission, super-galactic size scales (tens of kpc) of the line and
continuum emission, continuum spectral energy distribution similar to
those of present-day Magellanic Irregulars or very late type spiral
galaxies (but on average, redder than the nuclei of the associated
AGN), K magnitudes that fall along the band defined by the radio
galaxies in the IR Hubble diagram (K magnitude versus redshift), and
rest-frame optical/UV luminosities about ten times more luminous than
the most luminous galaxies in the present-day universe (\cite{hlvm91a};
\cite{hlmv91b}; \cite{lhcm92}; \cite{lvhm97a}).  Moreover, the
ground-based data revealed very asymmetric emission line and continuum
morphologies for the fuzz, and a weak tendency for quasars to exhibit
an ``alignment effect'' like radio galaxies (\cite{hlvm91a}).
Recent HST images now suggest that these asymmetric morphologies are
due to interactions with nearby companions, optical synchrotron
emission, scattering of light from the AGN, or to galaxies responsible
for intervening absorption systems along the line of sight to the
quasar.  They also imply that quasars exhibit a radio-UV/optical
alignment similar to radio galaxies (\cite{lehnert97b}).  In general,
the companion object or intervening absorber (which contributes a
substantial amount of the extended luminosity) does not lie along the
radio axis (\cite{lhcm92}; \cite{lvhm97a}) and tend
to weaken the ``alignment'' generally seen in the quasar population.

To further our understanding of the properties of the high redshift
quasar population, we obtained spectra of two high (z$\geq$2) redshift
quasars using the Keck 10m to obtain optical (rest frame UV) spectra of
a sample of high redshift radio loud quasar that are known to have
extended Ly$\alpha$ and continuum emitting ``fuzz'' around quasars
selected from \cite{hlvm91a}.  Generally, obtaining spectra of the
host galaxies of high redshift quasars will allow us to attack a number
of specific issues concerning the nature and properties of these
objects.  Specifically, with data like the ones presented here, we
would like to be able to address the issues of: What is the process
that is responsible for the ionization of the extended emission line
gas?  Is the gas photoionization by massive stars whose formation may
have been induced by the passage of the radio emitting plasma, shocks
due to the kinetic energy released by the AGN, photoionization by the
AGN, or something else?  What is the nature of the extended continuum
emission? Is the extended continuum scattered quasar light or due to
the stellar population of the host galaxy?  What are the kinematic
properties of the extended emission line gas? Is the gas likely to be
infalling?  Might we be seeing the formation of galaxy disks (some of
the nebulae are quite elongated)?  Is the gas interacting dynamically
with the radio emitting plasma?  Can we find evidence for rotation
which might suggest a more relaxed dynamical state, or find large scale
velocity shears that might implicate galaxy mergers as important in
fueling the ``quasar epoch''?  What is the relationship (if any)
between the z$_{abs}$ $\approx$ z$_{em}$ absorption line systems and
the extended emission line gas?

\section{Observations and Reduction}

The spectra were taken on September 1, 1995 (UT) using the Keck 10m
telescope in combination with the Low Resolution Imaging Spectrometer
(LRIS; \cite{oke95}) with a TEK 2048$^2$ CCD.  The night was cloudy and
we were only able to observe for a few of hours at the end of the night
when the clouds thinned sufficiently to warrant observing.  We used
LRIS in two configurations.  For the observations of \object{PKS
2338+042}, we used with the 600 l m$^{-1}$ grating (blazed at
5000\AA).  The resulting spectrum covered the spectral range 3860 --
6430\AA \ and had a resolution of about 4.5\AA \ (as measured using the
night sky lines).  This range was sufficiently large to include the
Ly$\alpha$, CIV$\lambda$1549, and HeII$\lambda$1640 emission lines.
The final spectrum of \object{PKS 2338+042} is a result of averaging
three separate exposures of 1800 seconds duration.  The slit for these
observations was oriented along PA=135$^\circ$ and was selected to be
along the most extended position angle of the Ly$\alpha$ emission as
revealed in the images of \cite{hlvm91a} and approximately along
the position angle of the most extended radio emission.  For the
observation of \object{PKS 0445+097}, we used the 300 l m$^{-1}$
grating (blazed at 7500\AA).  The resulting spectrum covered the
spectral range 3980 -- 9040\AA \ and had a resolution of 11\AA \ FWHM
(as measured using the night sky lines).  The spectral range was
sufficient to include CIV$\lambda$1549, HeII$\lambda$1640,
CIII]$\lambda$1909, and MgII$\lambda$2798.  We were only able to obtain
one 1800 second exposure of \object{PKS 0445+097} due to increasing
cloudiness.  The slit was oriented along PA=110$^\circ$ which was
chosen such that the slit would pass through the nucleus and the blob
of continuum emission seen in the HST images of \object{PKS 0445+097}
in \cite{lvhm97a}.  Both quasars were observed through a 1''
wide slit.  The scale of the CCD is 0.22'' pixel$^{-1}$.  The seeing
during the observations was about 1''.

The spectra were reduced in the standard way using IRAF.  The data were
bias subtracted using the values of the overscan region, flat-fielded
using several exposures of the internal lamp taken immediately after
the observations, and flux-calibrated (to remove the response only
since the night was not spectroscopic) using observations of Hz4.  The
spectra were wavelength calibrated using exposures of the Ne, Hg, and
Kr lamps.  The parameters of the lines (center, width, and flux) were
then estimated by fitting a Gaussian profile using the IRAF task
SPLOT.  The Ly$\alpha$ emission in the longslit spectrum of \object{PKS
2338+042} was obviously spatially extended.  We measured the properties
of the extended emission in sums of 4 CCD rows which is about 0.9'' in
projection on the sky.  Since the seeing was $\leq$1'', these
extractions provide independent information.  Error estimates were
provided by SPLOT assuming that the noise in each pixel was just the
root mean square of the noise in the surrounding continuum (see the
help for SPLOT).  The uncertainties in the fit estimated this way
probably slightly under-estimates the true uncertainty of the fit.
Generally, for the analysis of the extended line emission, we excluded
spatial regions of the spectrum that had strong contamination due to
line emission from the nucleus.  Thus none of our results rely on the
rather difficult and uncertain method of subtracting off a template
nuclear spectrum in order to reveal the narrow extended emission line
component.

Since the night was not photometric, we have compared our emission
line fluxes with those obtained through ground-based narrow-band
imaging from \cite{hlvm91a}.  We find that our measured
Ly$\alpha$ for \object{PKS 2338+042} is about a factor of 4 lower than measured
during photometric conditions by \cite{hlvm91a}.  Along similar
lines of reasoning, we have used the HST F555W (similar to ``V'') image
of \object{PKS 0445+097} to estimate the how much of the flux we may have missed
due to the cloudy weather.  We find little difference (within about
20\%) between the flux estimated from the HST image and that of the
Keck spectrum of \object{PKS 0445+097} (the clouds were patchy).

\section{Results}

\subsection{The Longslit Spectrum of \object{PKS 0445+097}}

\begin{figure}
\resizebox{\hsize}{!}{\includegraphics{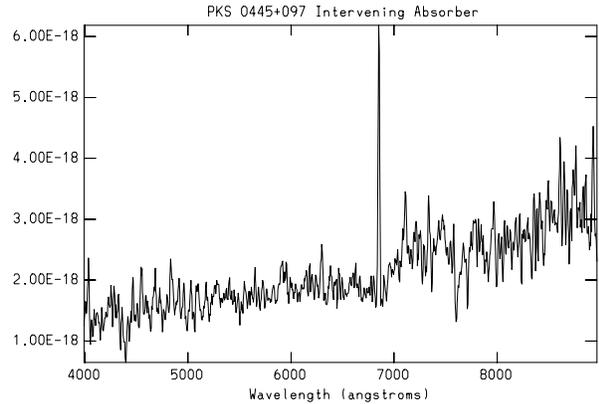}}
\caption{The spectrum of the intervening absorber 1.77'' from
the nucleus of \object{PKS 0445+097}.  This spectrum is a result of extracting
a region centered 1.6 arc seconds from the nucleus and over a region
1 arc second in diameter.  We have identified three lines in this
spectrum, [NeV]$\lambda$3426 at 6298.9\AA,
[OII]$\lambda$3727 at 6853.1\AA, and [NeIII]$\lambda$3869 at 7110.6\AA
which imply a redshift of 0.8384 $\pm$ 0.0002.  The identification and
strength of the [NeV]$\lambda$3426 line might be influenced by the
strong telluric line of [OI] at 6300\AA.}
\label{fig1}
\end{figure}

We oriented the slit along a position angle such that the spectrum
would obtain the redshift of the nearby (in projection) companion located
approximately 1.8'' from the quasar nucleus seen in the HST image of
\cite{lvhm97a}.  Using the nuclear HeII$\lambda$1640 and
CIII]$\lambda$1909 emission lines, we estimate the redshift of the
quasar to be 2.1083 $\pm$ 0.001.  The other strong lines in the
spectrum, namely, CIV$\lambda$1549 and MgII$\lambda$2800, had profiles
that were severely affected by absorption lines.  Inspecting the
extended emission of the quasar reveals that this nearby object to the
southeast of \object{PKS 0445+097} is an intervening absorber galaxy
(Fig.  \ref{fig1}).  The extracted spectrum reveals three identifiable
lines with central wavelengths of: 6298.9\AA, 6853.1\AA, and
7110.6\AA.  We note however that the line at 6298.9\AA is very close in
wavelength to the telluric [OI]$\lambda$ 6300 line.  Although the sky
subtraction appears to be good, it is possible that this line is
influenced by inadequacies in the night sky subtraction.  The lines are
detected at the 5$\sigma$, 30$\sigma$, and 8$\sigma$ level
respectively.  We identify these lines as [NeV]$\lambda$3426,
[OII]$\lambda$3727, and [NeIII]$\lambda$3869 which implies a redshift
of z=0.8384 $\pm$ 0.0002 (where the uncertainty in this determination
includes the random error from each individual line).  The lines are
not resolved at the low resolution of the spectrum.  From the nuclear
spectrum of \object{PKS 0445+097}, we identify a single absorption line
at 5147.1 $\pm$ 1\AA, which we identify (as \cite{btt90} did also) as
MgII$\lambda$2800 absorption at z=0.8396 $\pm$ 0.0004.  \cite{btt90},
who had a higher spectral resolution than the data presented here,
measured z=0.8392 for this absorption line system, which is within the
1$\sigma$ uncertainty of our measurement.  Calculating the velocity
difference between the absorption line system and the emission line of
the galaxy, we find a velocity offset of 200 $\pm$ 80 km s$^{-1}$.
Moreover, the relative fluxes in the lines give the following ratios
relative to [OII]:  f$_{[NeV]}$/f$_{[OII]}$$\approx$0.2 $\pm$ 0.2 and
f$_{[NeIII]}$/f$_{[OII]}$$\approx$0.3 $\pm$ 0.1.

\begin{figure}
\resizebox{\hsize}{!}{\includegraphics{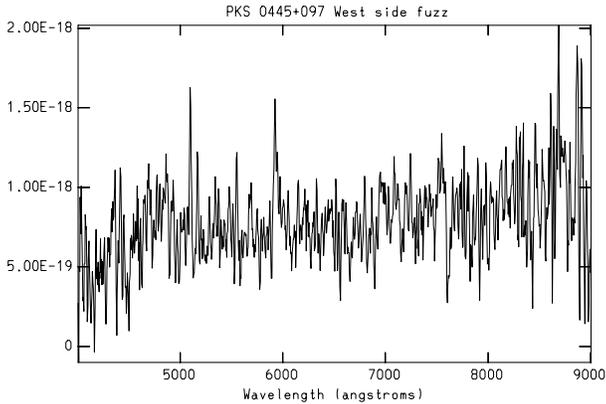}}
\caption{The spectrum of the extended emission from the quasar
\object{PKS 0445+097}.  The spectrum was extracted from a region that
spans from 1.4 arc seconds to 2.9 arc seconds from the nucleus of
\object{PKS 0445+097} and was specifically chosen to avoid the broad
line emission from the nucleus.}

\label{fig2}
\end{figure}

In addition to the flux associated with the intervening absorber, we
also see extended flux that is apparently associated with the quasar.
Because the southeastern side of the quasar is ``contaminated'' with
the emission from the intervening absorber, we have extracted a
spectrum whose center is about 2.3'' from the brightest quasar emission
and is a sum of about 7 CCD rows ($\approx$1.5'').  We display the
extracted spectrum of the fuzz on the northwest side of the nucleus in
Fig. \ref{fig2}.  The two obvious lines are HeII$\lambda$1640 and
CIII]$\lambda$1909.  Due to falling sensitivity of the chip, the bad
weather conditions and for the case of the MgII line, the strong night
sky emission, we were unable to obtain high signal-to-noise information
about the possibility of extended CIV$\lambda$1549 and
MgII$\lambda$2800.  We find that the HeII$\lambda$1640 line is
relatively narrow (FWHM=1000 $\pm$ 200 km s$^{-1}$) and that the profile
of the CIII]$\lambda$1909 line is much broader (FWHM=2200 $\pm$ 600 km
s$^{-1}$).  For the quasar nucleus, we find that the lines broader
still, FWHM(HeII$\lambda$1640)=2400 km s$^{-1}$ and
FWHM(CIII]$\lambda$1909)=5500 km s$^{-1}$.  The width of the extended
HeII and CIII] lines is comparable to the width of extended Ly$\alpha$
measured by \cite{hlmv91b} who measured about 1500 km s$^{-1}$
for the full width at half maximum.  The central wavelengths of the
extended HeII and CIII] imply a redshift of 2.1087 and 2.1058
respectively, in fair agreement with the redshift estimated from the
nuclear emission lines.  Moreover, we find that the ratio of
HeII$\lambda$1640/CIII]$\lambda$1909$\approx$0.55.  For comparison, a
rough estimate of the ratio for the quasar nucleus is
HeII$\lambda$1640/CIII]$\lambda$1909$\approx$0.13.  Estimating the
signal to noise near the location of the undetected CIV$\lambda$1549,
we find that the 3$\sigma$ upper limit to the strength of
CIV$\lambda$1549 gives CIV$\lambda$1549/CIII]$\lambda$1909$<$0.2.

\subsection{The Longslit Spectrum of \object{PKS 2338+042}}

\begin{figure}
\resizebox{\hsize}{!}{\includegraphics{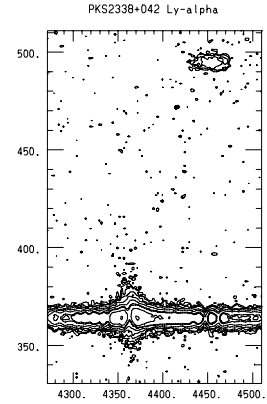}}
\caption{A section of the full 2-dimensional spectrum of \object{PKS
2338+042} shown to highlight the Ly$\alpha$ emission from both sources.
The bottom spectrum is of the quasar, while that at the top shows the
putative Ly$\alpha$ emission from the object serendipitously placed
along the slit.  The x-axis axis is in angstroms and the y-axis is in
pixels (0.22'' pixel$^{-1}$).  The displayed portion of the long
slit spectrum is approximately 240\AA \ $\times$ 40 arc seconds.}
\label{fig3}
\end{figure}

\begin{figure}
\resizebox{\hsize}{!}{\includegraphics{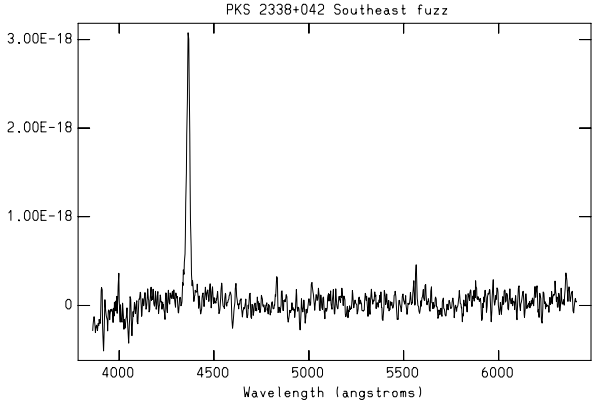}}
\caption{The top spectrum is that off the extended emission from
the quasar \object{PKS 2338+042} on the southeastern side of the nucleus.  The
bottom spectrum is that extracted from the extended emission on the
northwestern side of the nucleus along PA=135$^\circ$.  The two spectra
were extracted over a region that was 1.6 arc seconds to 7.5 arc
seconds on the southeastern side of the nucleus and 3.2 arc seconds to
6.7 arc seconds on the northwestern side of the nucleus.  These regions
were specifically chosen to avoid the obvious broad-line emission from
the quasar nucleus.  On the southeastern side of the nucleus, we
identify to weak lines other than Ly$\alpha$; CIV at 5565.3\AA \ and
HeII at 5880.7\AA.  On the northwestern side of the nucleus, we only
identify weak CIV at 5567.9\AA.}
\label{fig4}
\end{figure}

\begin{figure}
\resizebox{\hsize}{!}{\includegraphics{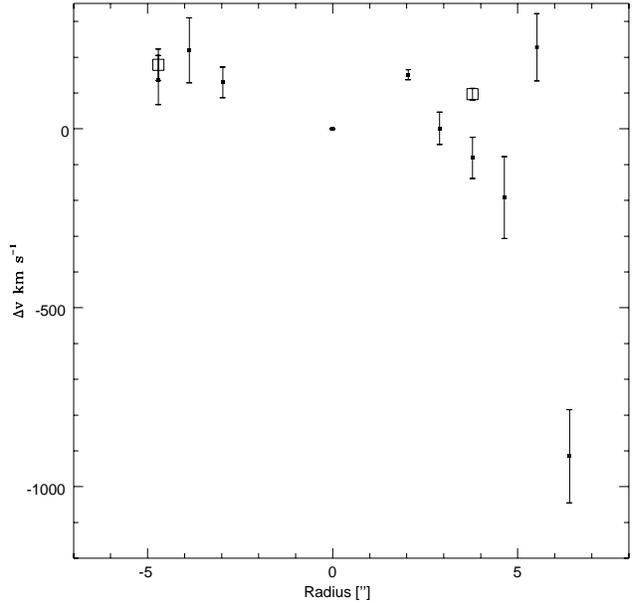}}
\caption{At the top, the spatially resolved velocity structure of
the extended Ly$\alpha$ emission from \object{PKS 2338+042}.  The zero-point of
the velocity scale was chosen to be the redshift of the strong
Ly$\alpha$ absorption line seen in the nuclear spectrum of \object{PKS
2338+042}.  The redshift of the Ly$\alpha$ absorption line is 2.5888,
which corresponds quite closely to that of the nucleus
HeII$\lambda$1640 line (i.e., 2.5889).  The bottom plot shows the
spatially resolved line widths of the extended Ly$\alpha$ emission from
\object{PKS 2338+042}.  The hollow squares in both plots represent the
integrated measurements on each side of the nucleus as measured on the
spectra shown in Fig.\ref{fig4}.}
\label{fig5}
\end{figure}

The longslit spectrum of \object{PKS 2338+042} reveals extended
Ly$\alpha$ emission.  As described previously, we extracted the
spatially-resolved Ly$\alpha$ emission in 0.9'' wide bins.  In
Fig.\ref{fig3}, we show a portion of the longslit spectrum specifically
selected to show the region around Ly$\alpha$.  As can be seen, the
emission along PA=135$^\circ$ is strongly one-sided and with the most
intense, most extended emission being on the side with the radio jet
and most intense, most distorted radio emission (see \cite{bmsl88}
and \cite{lvhm97a}).  In Fig.\ref{fig4}, we show a spectrum of the
sum of the extended emission from the southeastern and northwestern
sides of the nucleus along the slit of \object{PKS 2338+042}.  In
Fig.\ref{fig5}, we show the velocity structure of the Ly$\alpha$
emission.  We see that the velocity offsets of the extended Ly$\alpha$
emission relative to the nucleus of \object{PKS 2338+042} are
relatively small. We have chosen the wavelength of the absorption
(4362.9\AA) in the center of the Ly$\alpha$ emission from the quasar as
the zero-point for showing the velocity structure.  The wavelength of
the absorption corresponds to z=2.5888.  Unfortunately, it is difficult
to determine how the velocity of the Ly$\alpha$ absorption compares
with that of the emission since most of the observed lines are strongly
affected by associated absorption.  Only HeII has a profile that
appears to be relatively unaffected by strong associated absorption and
is relatively narrow (observed to be about 25\AA).  The observed
wavelength of the nuclear HeII line implies a redshift of 2.5889, in
close agreement with that of the associated Ly$\alpha$ absorption.  As
can be seen the overall flux weighted average velocity of the extended
emission line gas (represented by hollow squares in Fig.\ref{fig4}) is
small -- only being about 100 km s$^{-1}$ to the southeast and about
200 km s$^{-1}$ to the northwest.  Although, we should note that some
of the offsets are quite large in the very faint, most extended gas.
About 6.3'' to the southeast we measure a velocity offset of about 900
km s$^{-1}$.

The extended Ly$\alpha$ emission line is also very broad.  Overall the
extended Ly$\alpha$ emission has line widths of 1300 km s$^{-1}$ to the
southeast and about 1050 km s$^{-1}$ to the northwest.  There is some
tendency for the line widths to narrow with increasing angular distance
from the nucleus.  We also note in Fig.\ref{fig5} the width of the nucleus
Ly$\alpha$ absorption (indicated by the point at zero radius).  We can
see that the absorption width is very narrow (only 500 km s$^{-1}$
FWHM) compared to the width of the extended emission line gas (about
1300 km s$^{-1}$).  However, the measured width of the Ly$\alpha$
absorption is probably best described as a lower limit.  In making
this measurement, we assumed that the continuum against which the
absorption is taking place is accurately represented by the observed
peaks in the Ly$\alpha$ emission line profile.  Undoubtably, the profile
of the Ly$\alpha$ line is still increasing at the point of the highest
observed points of the line and thus the true amount of absorption
is probably much higher than we have estimated.

The individually extracted spectra that had significant Ly$\alpha$ did
not show significant emission from CIV$\lambda$1549 or
HeII$\lambda$1640.  However, in the sum of these spectra we do detect
very weak (but significant) CIV$\lambda$1549 and HeII$\lambda$1640
emission on each side of the quasar nucleus along the slit.  For the
spectrum on the southeastern side of the nucleus, the line identified
as CIV has a measured ratio of $\approx$0.06 relative to Ly$\alpha$
(the night was not photometric), a central wavelength of 5565.3\AA
\ (corresponds to z=2.593), a width of 17.2\AA, and is a 7$\sigma$
detection.  For HeII we measure f$_{HeII}$/f$_{Ly\alpha}$
$\approx$0.03, a central wavelength of 5880.7\AA \ (which corresponds
to z=2.586), a width of 10.7\AA, and the line is a $\sim$5$\sigma$
detection.  For the northwestern side of the nucleus we find that CIV
has a measured ratio of $\approx$0.15 relative to Ly$\alpha$, a central
wavelength of 5567.9\AA \ (corresponds to z=2.594), a width of 11.2\AA,
and is a 7$\sigma$ detection.  We only measure an upper limit to the
HeII emission on this side of the nucleus which implies
f$_{HeII}$/f$_{Ly\alpha}$ $<$ 0.06.

\begin{figure}
\resizebox{\hsize}{!}{\includegraphics{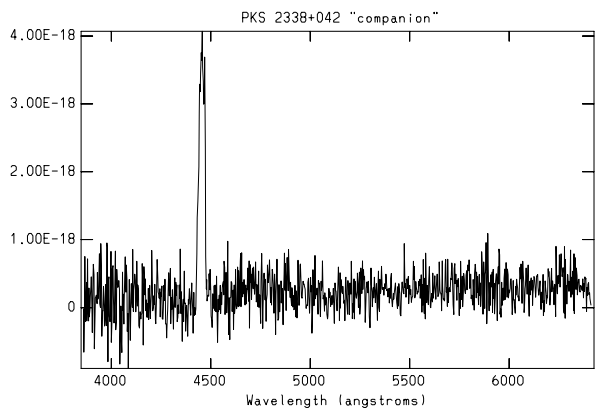}}
\caption{The spectrum of the ``companion'' galaxy that was
serendipitously placed along the slit.}
\label{fig6}
\end{figure}

In addition to seeing extended line emission from \object{PKS 2338+042}
we also see another object along the slit with a very strong line.
This object lies 29.2'' to the southeast of the nucleus of \object{PKS
2338+042} along PA=135$^\circ$.  We show the extracted spectrum in Fig.
\ref{fig6}.  We find that the one obvious line has a wavelength of
4455.0 $\pm$ 0.4\AA, a measured width of 32.0 $\pm$ 0.7\AA and a total
flux $>$1.3 $\times$10$^{-16}$ ergs cm$^{-2}$ s$^{-1}$.  The line is
also very broad (FWHM=2150 $\pm$ 50 km s$^{-1}$) and has a high
equivalent width (2100 $\pm$ 100\AA).  We have also measured the break
due the Ly$\alpha$ forest if this line is identified as Ly$\alpha$.  If
we quantify the amount of continuum depression due to Ly$\alpha$
absorption lines to the blue of the Ly$\alpha$ using the definition of
\cite{ssg91}, namely D$_A$=1-f$_\nu$(obs)/f$_\nu$(cont), where
f$_\nu$(obs) is the observed flux density at blueward of Ly$\alpha$ and
f$_\nu$(cont) is the flux density of the extrapolated continuum
blueward of Ly$\alpha$, we find that D$_A$=0.26 for the unidentified
object.  This is very similar to that measured for the quasar,
D$_A$=0.19, which is in turn consistent with measurements of quasars at
similarly high redshifts (see \cite{ssg91}).

\section{Discussion}

\subsection{Properties of the Extended Line Emission}

The properties of the extended Ly$\alpha$ emission in \object{PKS
2338+042} is typical of the high redshift quasar host observed
previously (see e.g., \cite{hlmv91b}).  Moreover, the properties of this
emission (flux, luminosity, width, equivalent width) are also very
similar to that observed for radio galaxies at similar redshifts (e.g.,
\cite{vojikphd}; \cite{mccarthy93} and references therein).  The line
ratios (and upper limit) observed however, are not typical of radio
galaxies.  High redshift radio galaxies typically have CIV and HeII
ratios relative to Ly$\alpha$ of $\approx$0.2 and 0.1 respectively.
However, the range of these line ratios is relatively large and there
are individual cases of radio galaxies which have CIV and HeII to
Ly$\alpha$ ratios as low as those observed here (e.g., 1138$-$262 and 4C
41.17, \cite{vojik97}; and see \cite{vmtc97}).  Therefore, although the
line ratios are not typical of radio galaxies, they certainly are
consistent with the range of line ratios observed in high redshift
radio galaxies.

In the extended emission from \object{PKS 0445+097}, we observe two
lines which we have identified as HeII$\lambda$1640 and
CIII]$\lambda$1909.  This confirms the possible detection of extended
HeII by \cite{hlmv91b} in \object{PKS 0445+097}.  The extended CIII] is
more surprising since that was not detected by \cite{hlmv91b}.
Although this is perhaps understandable since this region of the
spectrum was near the end of their spectral bandpass where the
sensitivity of the CCD was falling sharply.  The properties of the
extended CIII] are particularly interesting.  The width of the line is
significantly broader than that of the HeII emission.  Extended CIII]
emission is often observed in high redshift radio galaxies
(\cite{mccarthy93}) and references therein).  Typically it is about
50\% as strong as HeII as is observed for \object{PKS 0445+097}.
Moreover, it also appears that on average, the CIII] line is also
broader than HeII in the radio galaxies.  Since the critical density of
CIII] is high ($\sim$10$^9$ cm$^{-3}$) and the appropriate conditions
for excitation of strong CIII] are not generally seen in extended
narrow-line regions (e.g., \cite{docok89}) and since the ratio of
CIII]/HeII is much higher in the broad line emission than in the
extended emission, we wonder if the broad extended CIII] is related to
scattering of the nuclear light in both \object{PKS 0445+097} and radio
galaxies generally.  In the sample of \cite{vojikphd} there are 20
galaxies that have both HeII and CIII] measurements.  Of those 20 radio
galaxies, 11 have CIII] broader than HeII and the whole ensemble has a
ratio of CIII] to HeII widths of 1.35.  The breadth of the CIII] line
compared to the HeII line can also be gleaned from the composite high
redshift radio galaxy spectrum presented in \cite{mccarthy93}.

We conclude that for these two objects, the properties of their extended
line emission are consistent with similar results for high redshift
radio galaxies.

The source of the very large velocity dispersions observed in the
extended line emission is unknown.  If quasars are embedded in
cluster-like masses and the geometry is favorable, it is possible that
these large line widths are gravitational in origin (e.g., \cite{fr85};
\cite{hlmv91b}).  However, we consider such an explanation unlikely.  A
more likely explanation is that the powerful radio jets in these
quasars (\cite{lbm93}) is the source of power for generating the large
line widths.  There are several lines of evidence in support of this
hypothesis.  First is the anti-correlation between the radio size and
velocity dispersion found for high redshift radio galaxies found by
\cite{vojikphd}.  Within the context of this finding we note that both
\object{PKS 2338+042} and \object{PKS 0445+097} are both compact radio
sources (\cite{lbm93}) and we would thus expect them both to have large
velocity dispersions in their extended gas as they do.  The second is
the obvious ``jet-cloud interaction'' seen in some radio galaxies
(\cite{vojikphd}; \cite{cmv90}) and in the extended emission in some
quasars (\cite{lvhm97a}).  It is interesting to note within this
context that \cite{lvhm97a} find that \object{PKS 2338+042} exhibits a
particularly obvious ``jet-cloud interaction'' at the position where
the radio jet has its most severe bend (in projection).  Also, the
radio lobe nearest to the nucleus and the most distorted, most intense
radio emission in \object{PKS 2338+042} (\cite{lbm93}) also lies on the
same side of the nucleus as the brightest, most extended Ly$\alpha$
emission.  And finally is the observation that radio quiet quasars at
z$\approx$2 appear to have much lower extended Ly$\alpha$ luminosities
than radio-loud quasars suggesting a relationship between the radio
emission and exciting the extended line emission (\cite{lhl97}).  These
observations all suggest a significant relationship between the radio
emission and amount and kinematics of extended line emission.

\subsection{The Intervening Absorber Galaxy}

We have shown that the galaxy 1.77'' to the southwest of the quasar
nucleus of \object{PKS 0445+097} is at a redshift of 0.8384.  This
redshift closely corresponds to the redshift of the strongest MgII
absorption line seen in the spectrum of \object{PKS 0445+097}.  At this
redshift (with H$_0$=75 km s$^{-1}$ Mpc$^{-1}$ and q$_0$=0.0), the
separation on the sky of this galaxy from the line of sight of the
quasar is about 12.1 kpc.  This roughly consistent with the empirical
relationship between impact parameter and rest frame MgII equivalent
width found by \cite{lb90} The HST images reveal that this galaxy is a
very distorted system.  There is a bright compact ``nucleus'' with a
``trail'' of material to the southeast which itself has two high
surface brightness knots (see \cite{lvhm97a}).  The slit was oriented
such that it passed through the brightest of the knots of emission.
From the spectrum, we conclude from a comparison with the line ratios
and line widths of various types of emission line galaxies (starbursts,
radio galaxies, Seyfert 1s and 2s, LINERs) that this galaxy has a
Seyfert 2-like spectrum (see \cite{docok89}).

Of a more speculative nature.  \cite{blw95} have used the velocity
difference between a galaxy along the line of sight to two quasars and
an intervening low redshift Ly$\alpha$ absorption line due to the
galaxy along the line of sight to estimate the mass of the galaxy.
They assumed that the absorbing material was part of the galaxy
rotation curve.  In a similar spirit, we wish to estimate what the
redshift difference between the galaxy we have identified as giving
rise to the MgII absorption line in \object{PKS 0445+097} and that of
the absorption line itself implies about the mass of the intervening
galaxy.  We find that the redshift difference between the absorption
line and the galaxy is about 200 km s$^{-1}$.  The distance from the
absorbing galaxy from the line of sight to the quasar is about 12 kpc.
If we assume that the absorbing cloud is moving about the center of the
absorbing galaxy in a circular orbit, implies the mass of the system is
1.2$\times$10$^{11}$ M$_{\sun}$.  Since it is unlikely that the
absorbing material is in a circular orbit, this estimate is likely to
be an upper limit under the assumption that the material is
gravitationally bound to the galaxy.  Using the absolute (restframe) B
magnitude calculated from the HST image by \cite{lvhm97a}, we find a
mass to light ratio of this galaxy in the rest-frame B band to be about
8 (M$_{\sun}$/L$_{\sun}$)$_B$.  This mass-to-light ratio is high, but
still reasonable for a spiral galaxy.  Of course there is no evidence
that this material is gravitationally bound.  It may well be that in
fact this gas is infalling onto the disk or is outflowing due to
hydrodynamical processes with the galaxy (perhaps as an AGN or
starburst-driven wind; \cite{lh96}; \cite{baum93}; \cite{kb86}).
However, since MgII absorbers appear to be generally associated with
``normal'' galaxies (see \cite{sdp94}), it is perhaps unlikely that
this halo gas is associated with any spectacular process occurring
within this galaxy.

\subsection{The Nature of the Emission Line Object in the Spectrum of
\object{PKS 2338+042}}

With only one detected line in the spectrum of object fortuitously
along the slit of our integration on \object{PKS 2338+042} it is
difficult to argue definitively as to the nature of this object.  There
are two likely possible identification of this single line.  It is most
likely either to be Ly$\alpha$ at z=2.6646 $\pm$ 0.0005 or
[OII]$\lambda$3727 at z=0.1953 $\pm$ 0.0005.  There are several reasons
to believe that this line is Ly$\alpha$ emission at z=2.6646.

First the line is very broad with a full width at half maximum of about
2200 km s$^{-1}$.  Lines this broad have been observed for Ly$\alpha$
in high redshift radio galaxies (e.g., \cite{vojik97}) and the host
galaxies of radio loud quasars (e.g., \cite{hlmv91b}), but never in
galaxies with unambiguously identified [OII]$\lambda$3727.  Second, we
find that line has a very high rest-frame equivalent width.  The line
has a measured equivalent width of about 2100\AA.  This implies, if the
line is Ly$\alpha$, a rest-frame equivalent width of about 500\AA.  If
the line is [OII] then the line has a rest-frame equivalent width of
about 1800\AA.  Again, rest-frame equivalent width this high have been
observed for extended Ly$\alpha$ emission in high-z radio galaxies
(e.g.,  \cite{mccarthy93}), in the hosts of quasars (e.g., \cite{hlmv91b};
this study), and in some high redshift galaxies associated with damped
Ly$\alpha$ absorbers (e.g., \cite{macchetto93}), but is well outside
the distribution of [OII] equivalent widths ([OII] equivalent widths
$\leq$100\AA) measured for low-z star-forming galaxies (e.g.,
\cite{bes88}; \cite{colless}; \cite{kennicutt}).  And finally, we
measured a ``break strength'' across the line of this object that is
comparable to that of the quasar.  If this line were [OII] instead, it
is difficult to understand why we see a break at all across the line.

Identifying this line as Ly$\alpha$, we have obtained approximate upper
limits for the strength of CIV$\lambda$1549 and HeII$\lambda$1640,
assuming that both lines have the same width as the punitive Ly$\alpha$
line. This leads to limits in the ratio of f$_{CIV}$/f$_{Ly\alpha}<$
0.06 and f$_{HeII}$/f$_{Ly\alpha}<$ 0.07.  Comparing these upper limits
to the line ratios with that of AGN and star-forming regions, we find
that these limits are consistent with that expected of a HII region or
star-forming galaxy.  The upper-limits to the line ratios are
inconsistent with the ``mean spectrum'' for a Seyfert 2 galaxy given in
\cite{fo86}.  In spite of these rather mundane line ratios compared to
active galaxies implied by these upper limits, it is again difficult to
understand characteristics of this line within the context of
star-formation (see \cite{valls93}).  The observed flux of Ly$\alpha$
implies a total luminosity of $>$1h$^{-2}$$\times$10$^{43}$ ergs
s$^{-1}$ (which is likely to actually be about a factor of 4 higher
from the analysis given in \S 2; using H$_0$=75 km s$^{-1}$ Mpc$^{-1}$
and q$_0$=0.0).  The Ly$\alpha$ luminosity and equivalent width are
large compared to low redshift starburst galaxies (see
\cite{kinney93}), but are within the range observed for low-z active
galaxies (e.g., \cite{krk90}).

We note that several other groups have discovered Ly$\alpha$ emitters
near high redshift quasars (e.g., \cite{sds91}), radio galaxies, and
damped Ly$\alpha$ absorption systems (e.g., \cite{djorgovski96};
\cite{macchetto93}).  Many of the objects so discovered have properties
remarkably similar to the object discovered here.  One object in
particular is strikingly similar to the object discovered here, namely
G-Q0000-2619 from \cite{macchetto93}.  These authors have suggested
that the line emission is possibly due to star-formation.  They
estimate star-formation rate of about 20 M$_{\sun}$ \ yr$^{-1}$.  Since
this estimate is highly model dependent and are very sensitive to the
amount of extinction within this object (which is unknown), quoting
exact numbers for the star-formation rate makes little sense.  However,
for the purposes of comparison, the Ly$\alpha$ and UV continuum flux
observed in the object we have accidentally discovered are comparable
with the star-formation rate estimated in \cite{macchetto93} for
G-Q0000-2619.

The odd characteristics of this possible Ly$\alpha$ emitter makes us
wonder if it could be ionized by the radiation from the quasar
nucleus.  Using our spectrum of the quasar and extrapolating the
continuum to estimate the flux at 2500\AA \ in the rest-frame and then
using the relations in \cite{elvis} to estimate the number of
ionizing photons, we estimate that PKS2338+097 emits about
2$\times$10$^{56}$ ionizing photons s$^{-1}$.  We note that this
estimate is very uncertain and could be an order of magnitude higher or
lower.  Moreover, taking the size of the object to be its total extent
in Ly$\alpha$ as measured in the spectrum (about 3''), the distance to
be 29'' which corresponds to about 260h$^{-1}$ kpc, and that every
photon intercepted results in a Ly$\alpha$ photon.  The physical
distance we assume is a definite lower limit since it assumes that the
substantial difference in the measured redshifts (about 6000 km
s$^{-1}$) is not related to the Hubble expansion.  These assumptions
are perhaps conservative since it is unlikely that the projected
distance is the true distance (which will be substantially larger),
that the size we observe is the projected size the ionizing photons
from the quasar intercept, and since the covering fraction of neutral
gas in this object is unlikely to be one.  Using our somewhat
conservative assumptions, we find that the predicted luminosity of
Ly$\alpha$ would be about 2$\times$10$^{42}$ ergs s$^{-1}$.  This is
about an order of magnitude less than we observe.  The number of
ionizing photons may be up to a factor of about 10 larger, but it might
also be lower.  In spite of the uncertainties, it seems very unlikely
that the quasar could provide enough ionizing photons to excite the
line emission we see, especially given that it is highly probable that
the quasar and emission line object are much more widely separated than
we have assumed here.

Given that we have only limited information as to the nature of this
object, we consider its exact nature very uncertain.  However,
the large FWHM and high equivalent width of the line we have identified
as Ly$\alpha$, we consider it likely that this object is a
high redshift AGN.

\section{Summary and Conclusions}

We present spectra taken with the Low Resolution Imaging Spectrograph
on the Keck 10m telescope of spatially-resolved structures (\lq
fuzz\rq) around the high-redshift radio-loud quasars \object{PKS
0445+097} (z=2.110) and \object{PKS 2338+042} (z=2.594).  For
\object{PKS 0445+097} we oriented the slit such that it passed through
the quasar nucleus and through an object identified on an HST exposure
1.8'' to the southeast of the nucleus.  We find that this object is at
a redshift of 0.8395 and has line ratios suggesting it has a Seyfert 2
nucleus.  The redshift of this object is very close ($\Delta$v=200 km
s$^{-1}$) to that of a MgII absorption line identified in the spectrum
of \object{PKS 0445+097}.  Moreover, we find that there is faint
extended HeII$\lambda$1640 and CIII]$\lambda$1909 from the quasar host
galaxy. Both of the lines are broad (FWHM$_{HeII}$=1000 $\pm$ 200 km
s$^{-1}$ and FWHM=2200 $\pm$ 600 kms$^{-1}$).  The limits on the
fluxes, the large widths, and the line ratio HeII/CIII] are similar to
what is observed for high redshift radio galaxies.

From the spectrum of \object{PKS 2338+042}, we find that Ly$\alpha$ emission
is extended up to 10'' from the quasar nucleus and is strongly one
sided (on the side of the stronger, closer to the nucleus, more
distorted radio jet and lobe).  The extended Ly$\alpha$ line is broad
(FWHM $>$ 1000 km s$^{-1}$) and the extended emission shows no evidence
for a strong velocity shear.  We also find weak extended CIV and HeII
emission which has similar characteristics (line widths, line fluxes,
and line ratios) to that of high redshift radio galaxies.  In addition,
we serendipitously discoved what appears to be a Ly$\alpha$ emitting
galaxy about 29'' to the southeast of the quasar at z=2.665.  The
single line identified in the spectrum has a measure equivalent width
of 2100\AA \ and a full width at half maximum of 2150\AA.  The high
equivalent width and large width of the line suggest that this galaxy
must be a high redshift AGN, although we note that the upper limits on
the CIV and HeII line ratios are mildly inconsistent with this galaxy
harboring an AGN.  A crude calculation shows the quasar nucleus is not
exciting the Ly$\alpha$ emission from this this galaxy.

\begin{acknowledgement}
The work of MDL at IGPP/LLNL was performed under the auspices of the US
Department of Energy under contract W-7405-ENG-48 and while at Leiden,
this work was supported by a program funded by the Dutch Organization
for Research (NWO).  The authors would like to thank the referee, Tim
Heckman, for his insightful comments that substantial improved this paper.
\end{acknowledgement}

{}

\end{document}